\documentclass[review]{elsarticle}

\usepackage{amsmath}
\usepackage{lineno,hyperref}
\modulolinenumbers[5]

\journal{Geographical Analysis (GA)
}

\usepackage{numcompress}\biboptions{authoryear}

\begin{document}

\begin{frontmatter}

\title{Effects of Aggregation Methodology on Uncertain Spatiotemporal Data}
%\tnoteref{mytitlenote}}
%\tnotetext[mytitlenote]{INSERT TITLE NOTE IF NEEDED}

\author{Zachary T. Hornberger\fnref{first}} 
\author{Bruce A. Cox \corref{mycorrespondingauthor}\fnref{second}}
\author{Raymond R. Hill \fnref{third}}
\address{Air Force Institute of Technology, 2950 Hobson Way, Wright Patterson Air Force Base, Ohio, 45433}

\fntext[first]{Zachary.Hornberger.2@us.af.mil}
\fntext[second]{BruceACox1@gmail.com}
\fntext[third]{RayRHill@gmail.com}
\cortext[mycorrespondingauthor]{Corresponding author, BruceACox1@gmail.com}

\begin{abstract}
Large spatiotemporal demand datasets can prove intractable for location optimization problems, motivating the need to aggregate such data.  However, demand aggregation introduces error which impacts the results of the location study. We introduce and apply a framework for comparing both deterministic and stochastic aggregation methods using distance-based and volume-based aggregation error metrics.  In addition we introduce and apply weighted versions of these metrics to account for the reality that demand events are non-homogeneous.  These metrics are applied to a large, highly variable, spatiotemporal demand dataset of search and rescue events in the Pacific ocean.  Comparisons with these metrics between six quadrat aggregations of varying scales and two zonal distribution models using hierarchical clustering is conducted. We show that as quadrat fidelity increases the distance-based aggregation error decreases, while the two deliberate zonal approaches further reduce this error while utilizing fewer zones. However, the higher fidelity aggregations have a detrimental effect on volume error.  In addition, by splitting the search and rescue dataset into a training and test set we show that stochastic aggregation of this highly variable spatiotemporal demand appears to be effective at simulating actual future demands.
\end{abstract}


\end{frontmatter}

%\linenumbers
%\end{document}
\section{Introduction}

\par Location modeling is a branch of operations research with vast real-world applicability and thus has been studied for a number of decades.  Location modeling typically considers the location and time of demand signals over a network and optimizes the corresponding location of a servicing asset, such as a factory or vehicle. The underlying spatiotemporal demand signal data points are thus instrumental to the quality of the resulting model.  

\par When considering large spatiotemporal datasets, there is frequently a need to aggregate demand points to make the problem more tractable for the solver, clearer for the analyst, and comprehensible for the end-user.  Aggregation, while of practical use, is not a lossless compression, and introduces aggregation error into the model.  When location data is aggregated, the resulting grouping's location is traditionally represented by an aggregated data point.  The distances between the actual demand points and the aggregated data points depend on the size of the aggregated region and the manner of aggregation. Similarly, the magnitude of uncertainty in the aggregated demand volumes is influenced by the nature of the aggregation.  Therefore, great consideration must be given to the aggregation technique used when solving location problems.

\par The impact of aggregation becomes more pronounced when the geographic region expands in size and there is high variability in demand density across the region; this struggle is actualized in studying the United State Coast Guard (USCG) District 14's search and rescue (SAR) mission.  The international community recognizes the need for global cooperation in responding to emerging crises around the world.  Nations have entered into SAR agreements, dividing the globe into respective search and rescue regions (SSRs).  Per the United States National Search and Rescue Supplement to the International Aeronautical and Maritime Search and Rescue Manual \citep{NatSAR}, the USCG is the federal SAR coordinator for SAR missions within the United States' maritime SSRs and the aeronautical SSRs that do not overlay the continental United States or Alaska.

\par USCG District 14 is headquartered in Honolulu, Hawaii and is responsible for USCG statutory missions across the Pacific region.  In particular, the district's SSR spans more than 12 million square nautical miles, though the preponderance of SAR emergencies occur in the vicinity of Guam and the Hawaiian Islands.  Additionally, District 14 has among the fewest assets in the USCG fleet, increasing the necessity to optimally posture those assets across the Pacific.  Given the time-sensitive nature of rescue operations, it is imperative the USCG be optimally postured to ensure rapid response.  Over the past decade, researchers have partnered with Coast Guard units - USCG and international - to solve these variations of the traditional facility location problem.  These studies typically use historic SAR event data as the foundation of either a deterministic or simulation-based location model.

\par This study quantifies the effects of the aggregation trade-off for spatiotemporal data over a large region, using District 14 SAR emergency data as a practical basis for consideration.  Section 2 of this paper reviews previous works related to the aggregation of data for location models in general and coast guard SAR missions in particular.  In section 3, we outline the methodology for implementing various aggregation techniques, both deterministic and stochastic, using a training data set.  In section 4, we evaluate the effectiveness of these techniques by quantifying the aggregation errors between the modelled demand and actual demand over a two-year period.  In section 5, we review our findings and provide recommendations for future research.

\section{Related Works}

\par Researchers have long been cognizant of a relationship between the methods used to aggregate location data and the resulting solutions generated by location models using this data.  \cite{Gehlke1934} were among the first to note this problem, observing that the smoothing of census data inherent in aggregation resulted in a loss of valuable information and impacted the corresponding correlation coefficients of their models.  \cite{Hillsman1978} laid a foundation for aggregation theory when they classified three sources of error (type A, B, and C) associated with representing individual demand points using aggregated demand points for solving factory location problems.  Source A refers to the difference in distance from the aggregated demand points to the placed factory and the sum of distances from individual demand points to the factory.  Source B is similar to Source A, if the factory were required to be collocated with an aggregated demand point.  Source C refers to the phenomena where individual demand points are erroneously assigned to inefficient factories due to the zone in which it is aggregated.

\par Several research teams have subsequently sought to quantify and minimize these aggregation errors.  \cite{Papadimitriou1981} presents two heuristics for aggregating data points in a manner that reduces the worst-case aggregation error and \cite{Zemel1984} produced a theorem for the worst-case bounds on Papadimitriou's honeycomb approach.  \cite{Qi2010} note the underlying assumption to Zemel's work of uniformly distributed demand points, and propose a multi-pattern tiling approach for considering arbitrarily distributed demand.  Works by \cite{Current1987, Current1990}, outline methods for eliminating Source A and B error when solving P-Median, set covering, and maximal covering location problems.  \cite{Francis2014} present a metric for measuring the error bounds for a P-Median problem, and \cite{Francis2004b} discuss formulations for minimizing the aggregation error using a penalty function approach.

\par
\begin{figure}[h]
\centering
\includegraphics[width=0.85\textwidth]{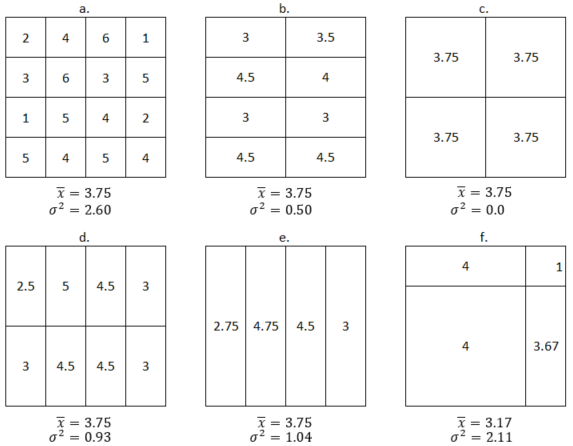}
\caption{(a-c) show effect of scale effect.  As scale of aggregation increases mean does not change but variance declines.  (d-e) show effect of aggregation effect.  Keeping scale equal but changing method of aggregation changes variance.  (c,e,f) show that even when number of zones is constant (4) mean and variance can change. }
\label{JelWU97}
\end{figure}

 In the fields of geography and ecology, aggregation error of spatial data points is dubbed the modifiable areal unit problem (MAUP) \citep{Openshaw, Dark2007} or the zone definition problem \citep{Fotheringham1995, Curtis1996}.  Research into MAUP typically decomposes the problem into two main effects: the scale effect and the zone effect.  The scale effect refers to the impact on the spatial analysis results that are caused by the fidelty of the aggregation; for example, the impact of aggregating demand in a city using 200m x 200m grids versus 1km x 1km grids.  Conversely, the zone effect refers to the impact caused by the way in which aggregation zones are bounded; for example, the impact of aggregating demand in a state using county lines versus city limits versus a grid overlay \citep{Openshaw, Dark2007}.   \cite{Jelinsky&Wu} created a seminal contrived demonstration of these effects, which we replicate for completeness in Figure \ref{JelWU97}.  

\par Previous research on MAUP has cautioned against arbitrary aggregation of spatial data and stressed its threat on the reliability of the resulting location analysis.  \cite{Openshaw} was foundational in the study of MAUP and called for developing better methods for aggregating spatial data due to MAUP's impact on the reliability of geographic studies.  \cite{Curtis1996} studied data for New York and concluded that researchers can bias the results of their analysis based on the means of aggregation, even if there appears to be a logical basis for the employed method of aggregation.  \cite{Fotheringham1995} go so far as to question the accuracy of any location-based analysis conducted using aggregated data because of the effects of MAUP.

\par In studies of MAUP, and aggregation theory in general, trends have emerged.  Increases in the number of aggregated zones are typically proportional to decreases in distance-based aggregation error; distance-based aggregation error disappears when each distinct demand point is assigned to a unique zone (i.e., the number of aggregation zones equals the number of demand points).  As any grouping introduces an associated level of distance-based error, it follows that reducing the amount of aggregation would subsequently reduce this error.  \cite{Francis2004} notes the law of diminishing returns applies in this context, however, suggesting that iterative reductions in the number of aggregate groups shows diminishing improvements to error reduction.  \cite{Francis1992} discuss the \textit{paradox of aggregation}, noting that solving formulations to minimize error can be more cumbersome than the original location problem being solved, which is counter-intuitive as aggregation is employed to simplify the resolution of these original location problems.  \cite{Dark2007} consider the trends corresponding to both the scale effect and the zone effect.  A known benefit of aggregation is tied to the scale effect; predictions of aggregated demand levels tend to be more accurate with fewer, larger aggregate zones.  This is because when there are more demand points consolidated in each zone, the demand variance between zones decreases.  The impact of zone effect is less understood and tends to differ from problem-to-problem.

\par The importance of careful aggregation has been thoroughly studied and is synthesized by \cite{Francis2008}.  In their survey of previous literature regarding aggregation error associated with location problems, Francis et al. note that there is an inherent tradeoff when aggregating data points; although aggregation has a tendency to decrease computational requirements and statistical uncertainty within the grouped data, it increases the error within the model by introducing aggregation error.  Thus there does not exist a singular ``best'' level of aggregation and the tradeoffs inherent in aggregation must be considered.

\par In addition to the theoretical work on this problem, there has been applied work specifically relating to Coast Guard SAR missions.  Although some research into this area was conducted in the late 1970s \citep{Armstrong}, the preponderance of studies relating to Coast Guard posturing has emerged in the past decade.  Studies researching the allocation of SAR assets, or facilities, typically adopt a quadrat modeling technique for aggregating location data \citep{AkbariINFOR, Akbari, Karatas, Afshartous}.  This technique consists of decomposing the region in question into square cells using a grid overlay.  Notably, the quadrat method is frequently adopted in crime data analyses, which typically seek to quantify spatial trends in criminal activity across a city or state \citep{Anselin, Chainey}.

\par \cite{Armstrong} constructed a goal-programming model for assigning SAR aircraft, incorporating probabilistic consideration for the time required by the aircraft to locate distress events in different areas of the corresponding region, using a grid overlay to create a collection of square zones.  These zones were then assigned deterministic values, representing the average number of distress events per month.  Similarly, \cite{Karatas} utilized a quadrat model for simulating the location and volume of distress calls for the Turkish Coast Guard in the Aegean Sea.  They first determined the optimal resource allocation strategy using individual events as separate demand nodes, and then evaluated the effectiveness of this strategy using simulated demand. 

\par The incorporation of kernel density estimation with the quadrat model, popular in crime data analysis \citep{Anselin}, has been previously implemented in SAR location problems.  The kernel density estimation method composes the region into grid cells and assigns a density function to each data point ($s_i$).  Points that are within proximity to each other relative to a specified bandwidth ($\tau$), are grouped into a kernel ($k$) and their density functions are combined.  The resulting image is a smooth heat map with greater densities illustrated over areas that have the most activity clustered closely together \citep{Anselin, Chainey}.  \cite{Erdemir} utilized kernel density estimation when considering the problem of locating aeromedical bases across the state of New Mexico.  Similarly, \cite{Akbari} implemented a kernel density estimation approach to approximate the intensity of distress calls received by the Canadian Coast Guard.  They varied the size of the grid overlay based upon the proximity to shoreline.  This decision was based upon the assumption that since most distress events occurred closer to shore, the analysis would benefit from greater fidelity in aggregation along the coastline.

\par Though not specifically kernel density estimation, \cite{Afshartous} implemented an intensity function-based approach for solving the Coast Guard SAR location problem.  They first constructed a non-parametric statistical simulation of distress calls within USCG District 7 (headquartered in Miami, Florida) and then utilized their simulation to model demand for a facility location problem.  This simulation was constructed by overlaying the region with a relatively fine grid and estimating the intensity of distress calls for each cell.

\par While most work regarding SAR posturing has incorporated quadrat techniques, \cite{Azofra} introduced an intuitive method that has been applied to maritime research.  Instead of defaulting to grids, Azofra's zonal distribution model allows for flexibility in the definition of emergency zones, such as zones based upon subject matter expertise.  Once the zones are determined, the centroids of distress calls, dubbed \textit{superaccidents}, are computed for each zone.  The zonal distribution model is a gravitational model, with the determination in optimal SAR operational response based upon the distance to the superaccidents and their associated weight.  They demonstrate the implementation of this model using a notional example involving three superaccidents and three ports.

\par Since the introduction of the zonal distribution model, some researchers have opted to expand upon it by applying it to real-world problems.  \cite{Ai} utilize this model for locating supply bases and positioning vessels for maritime emergencies for a portion of the coastline of China along the Yellow Sea.  While not adhering to the strict grid cells of previous studies, their zones remained rectangular in shape and varied in size across the region.  \cite{Razi} improved upon the zonal distribution model by utilizing a \textit{k}-means clustering algorithm for defining the zones and implementing a weighted approach for locating the superaccidents.  By adopting this approach, Razi et al. define the aggregated zones and corresponding representative demand nodes based upon historical trends in distress calls in the Aegean Sea rather than arbitrary cells.  \cite{Hornberger} propose an extension to the work of Razi and Karatas, which they dub the stochastic zonal distribution model.  Their model implements hierarchical \textit{k}-means clustering algorithm to define the aggregation zones, fits probability distributions to model the SAR demand for each zone, and then uses empirically constructed discrete distributions to model the corresponding rescue response for each emergency. 

\par A review of the existing literature regarding SAR asset posturing models finds a lack of explicit consideration regarding the impact of aggregation.  Additionally, as SAR research expands to larger regions of consideration (e.g., oceans vs. seas or shorelines), it is necessary to more thoroughly consider the effects of various aggregation methods. Outside of SAR, and more generally emergency response asset modeling (e.g., \cite{araz2007fuzzy}), other transportation resource posturing problems which utilize massive demand data-sets assume or require demand aggregation (e.g., taxi service areas (\cite{li2019taxi}, \cite{rajendran2019insights}), and should also be concerned with how such aggregation effects the associated location modeling.  To provide such consideration, our study utilizes historic SAR data from across the Pacific Ocean to compare the effectiveness of a zonal aggregation technique compared to quadrats of varying fidelity.  Additionally, we evaluate these tradeoffs in the aggregation as applied to deterministic and stochastic implementations.

\section{Methodology}

\par In this section, we consider two key characteristics that define a zonal aggregation of demand signals: dividing the region into zones, and modeling the demand level.  Using these two characteristics as the framework, we model and compare the following methodologies: deterministic quadrat approaches of various fidelities, the \cite{Razi} zonal distribution model, and the \cite{Hornberger} stochastic zonal distribution model.  

\par
\begin{figure}[h!]
\centering
\includegraphics[width=0.85\textwidth]{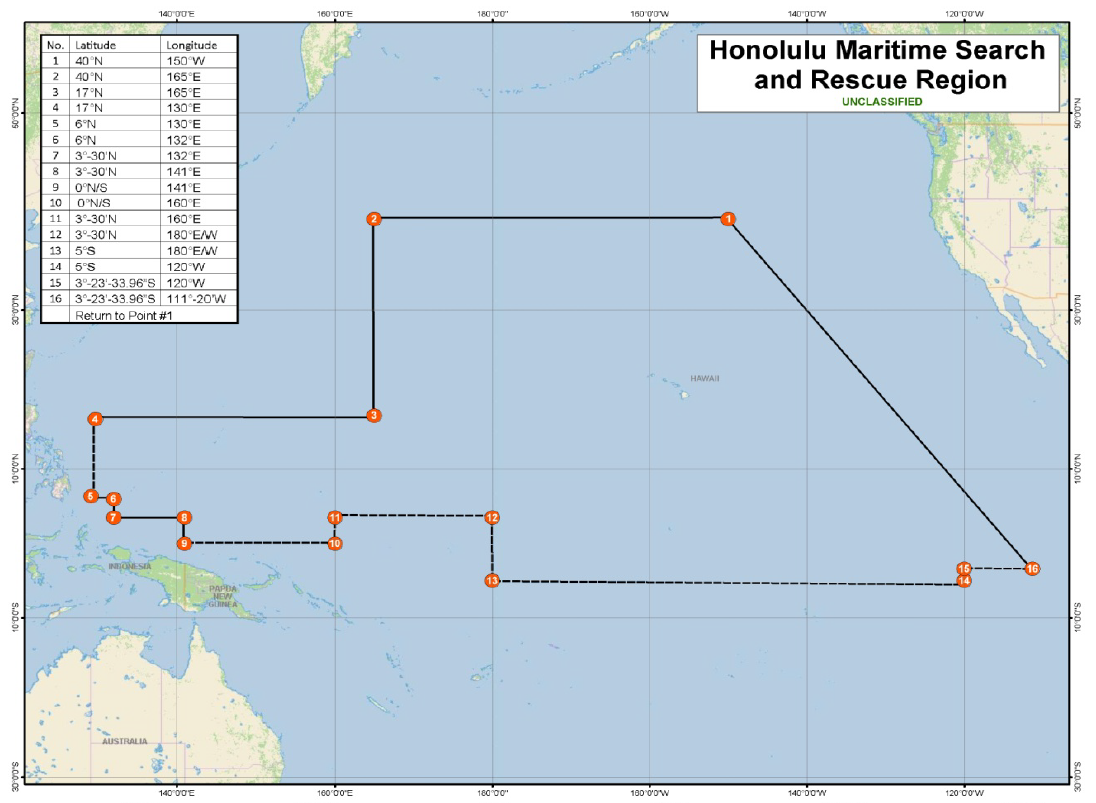}
\caption{Honolulu Maritime SAR Region}
\label{SAR_Region}
\end{figure}

 These methodologies are compared using the District 14 SAR region, an interesting test case due to its large area and highly variable demand levels; Figure \ref{SAR_Region} depicts the Honolulu Maritime Search and Rescue Region \citep{SARPlan}.  Historic search and rescue demand data was obtained from the Marine Information for Safety and Law Enforcement (MISLE) database to form both a training set and a test set.  The training set is comprised of SAR events from a 5 year span (January 2011 - December 2015) and is utilized to construct the models of spatiotemporal SAR demand.  The accuracy of the aggregated demand methodologies is then evaluated using historic SAR data for the same region from January 2016 - December 2017.

The training and test data is scoped to only consider events that occurred within the District 14 area of responsibility (AOR).  Additionally, demand points missing GPS coordinates were removed as were data points classified as medical consultations since these consultations only require a discussion with a medical professional over the phone and resources are not dispatched.  The final training set contains 2629 demand points and the test set contains 1080 demand points.

\subsection{Modeling Spatiotemporal Demand}

\par
\begin{figure}[h!]
\centering
\includegraphics[width=0.85\textwidth]{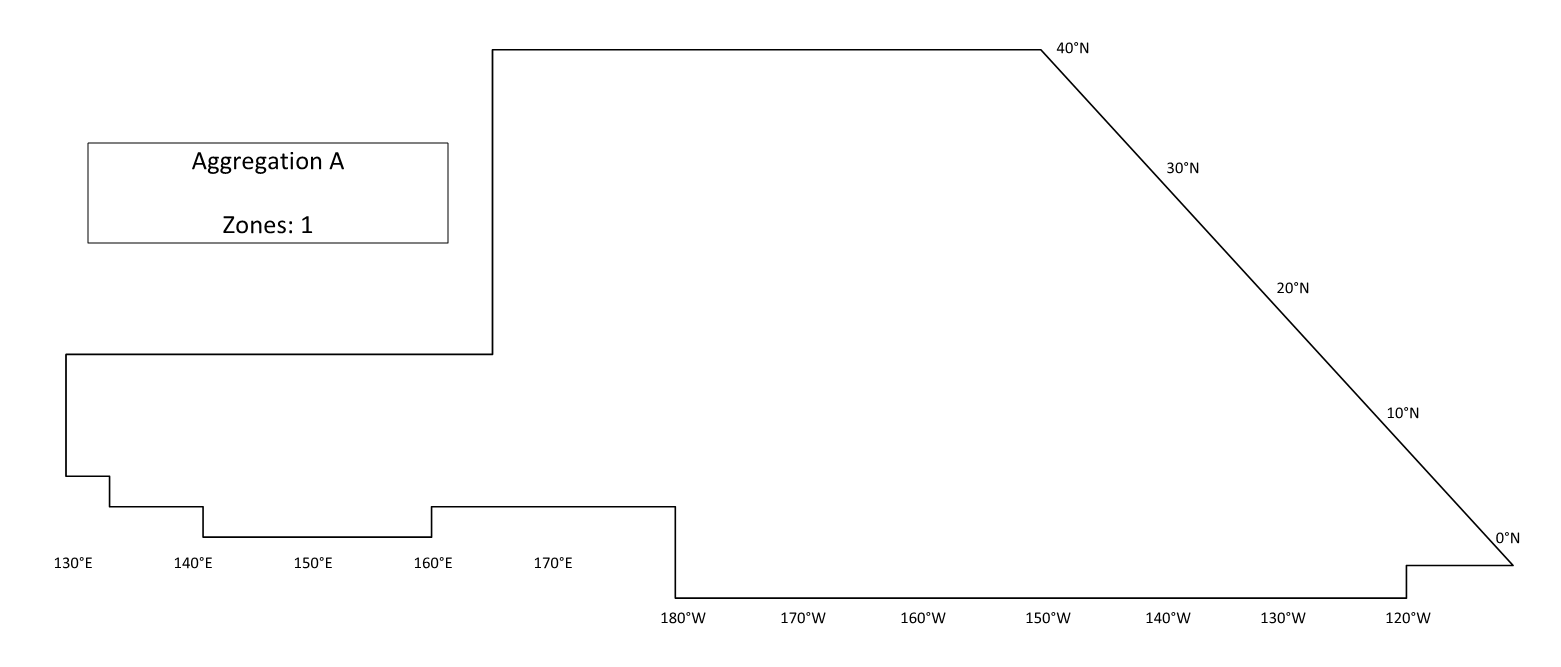}
\caption{Aggregation A (1 Zone)}
\label{AggA}
\end{figure}

\begin{figure}[h!]
\centering
\includegraphics[width=0.85\textwidth]{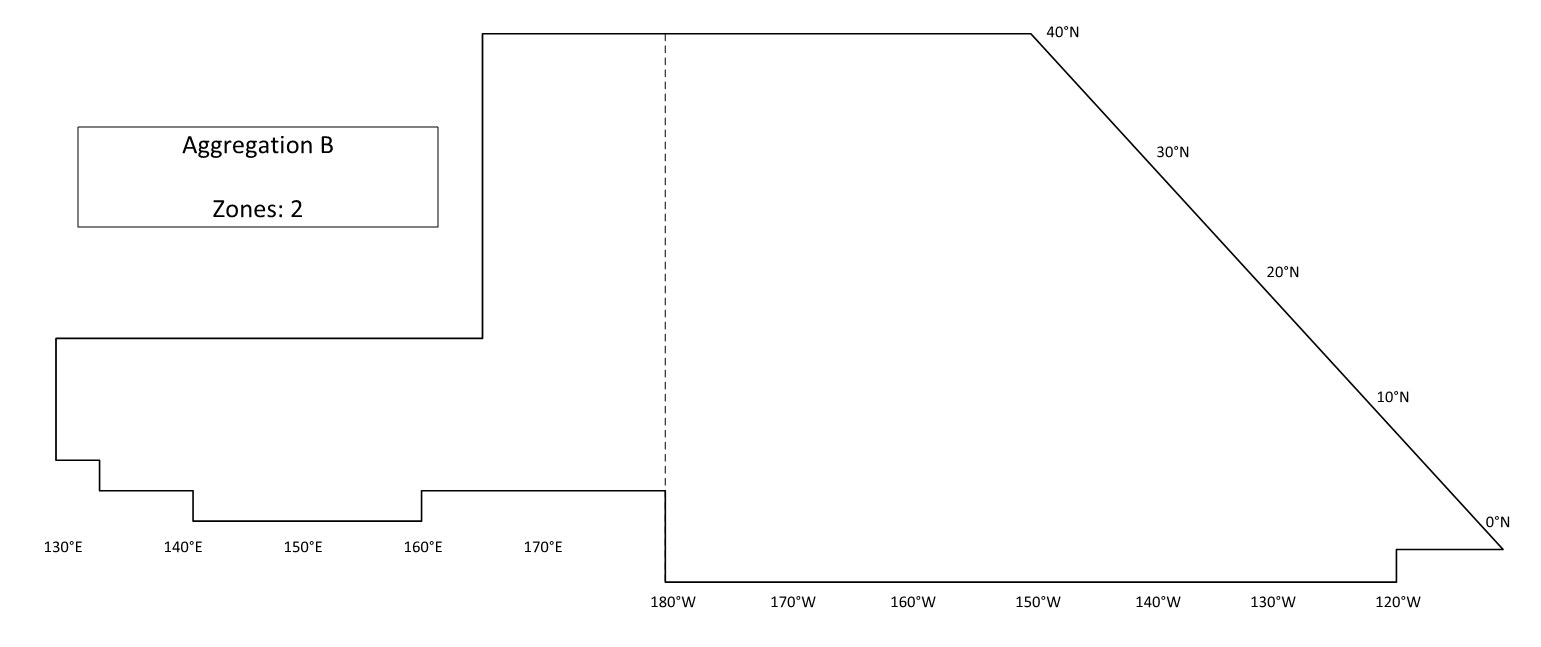}
\caption{Aggregation B (2 Zones)}
\label{AggB}
\end{figure}

\begin{figure}[h!]
\centering
\includegraphics[width=0.85\textwidth]{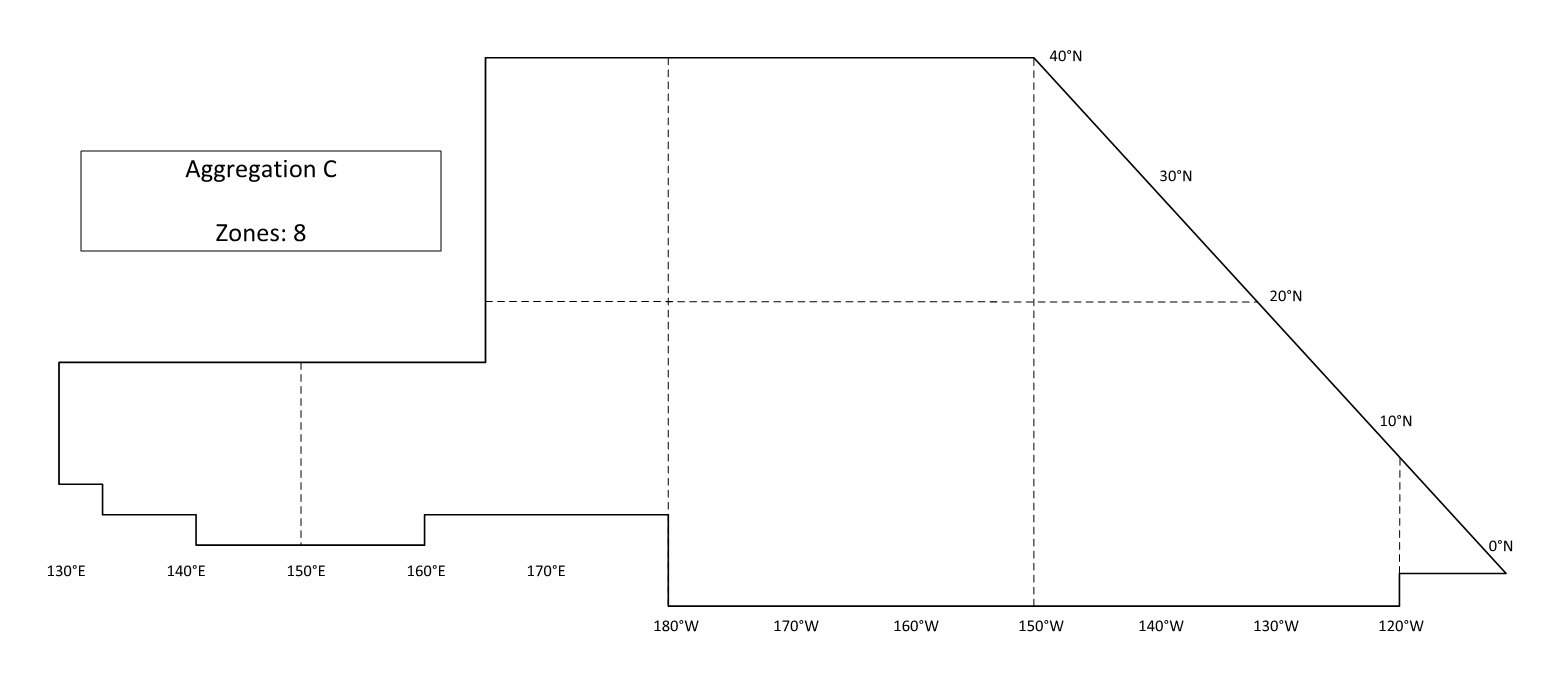}
\caption{Aggregation C (8 Zones)}
\label{AggC}
\end{figure}

\begin{figure}[h!]
\centering
\includegraphics[width=0.85\textwidth]{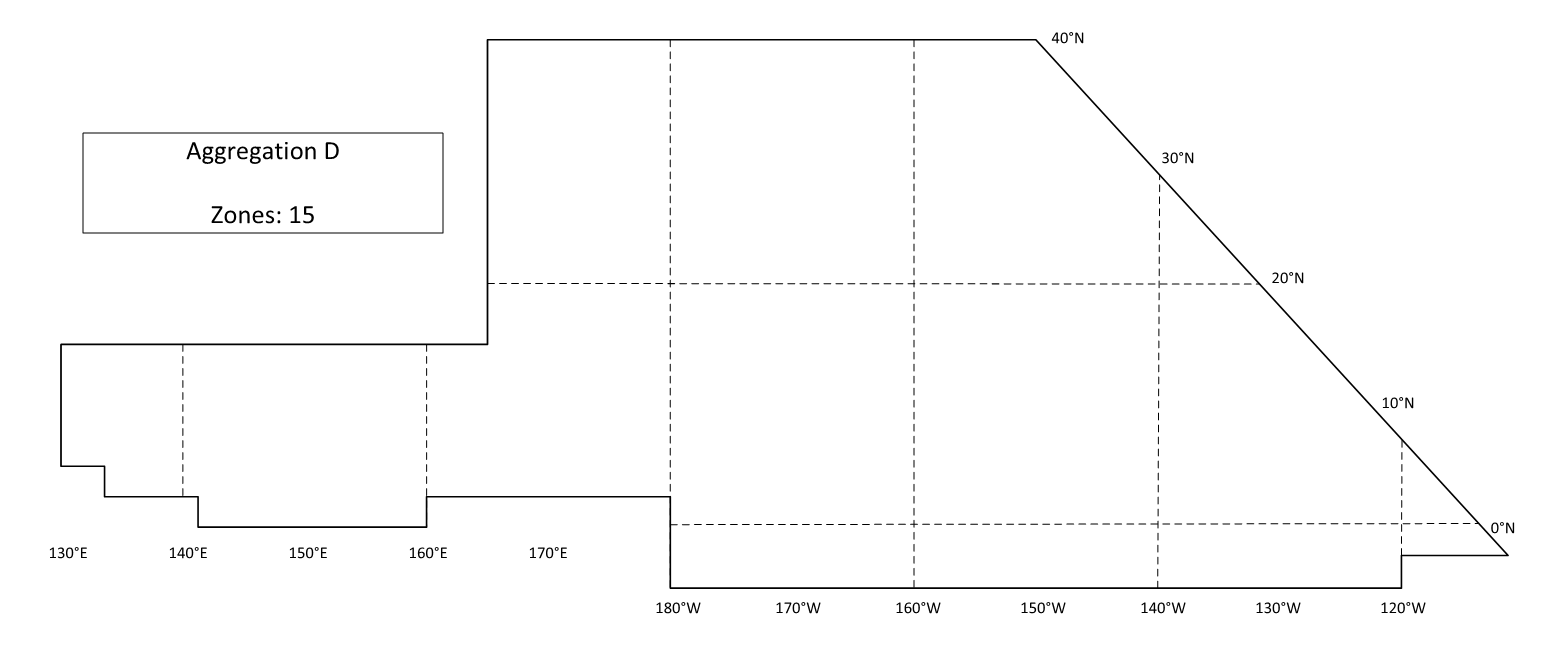}
\caption{Aggregation D (15 Zones)}
\label{AggD}
\end{figure}

\begin{figure}[h!]
\centering
\includegraphics[width=0.85\textwidth]{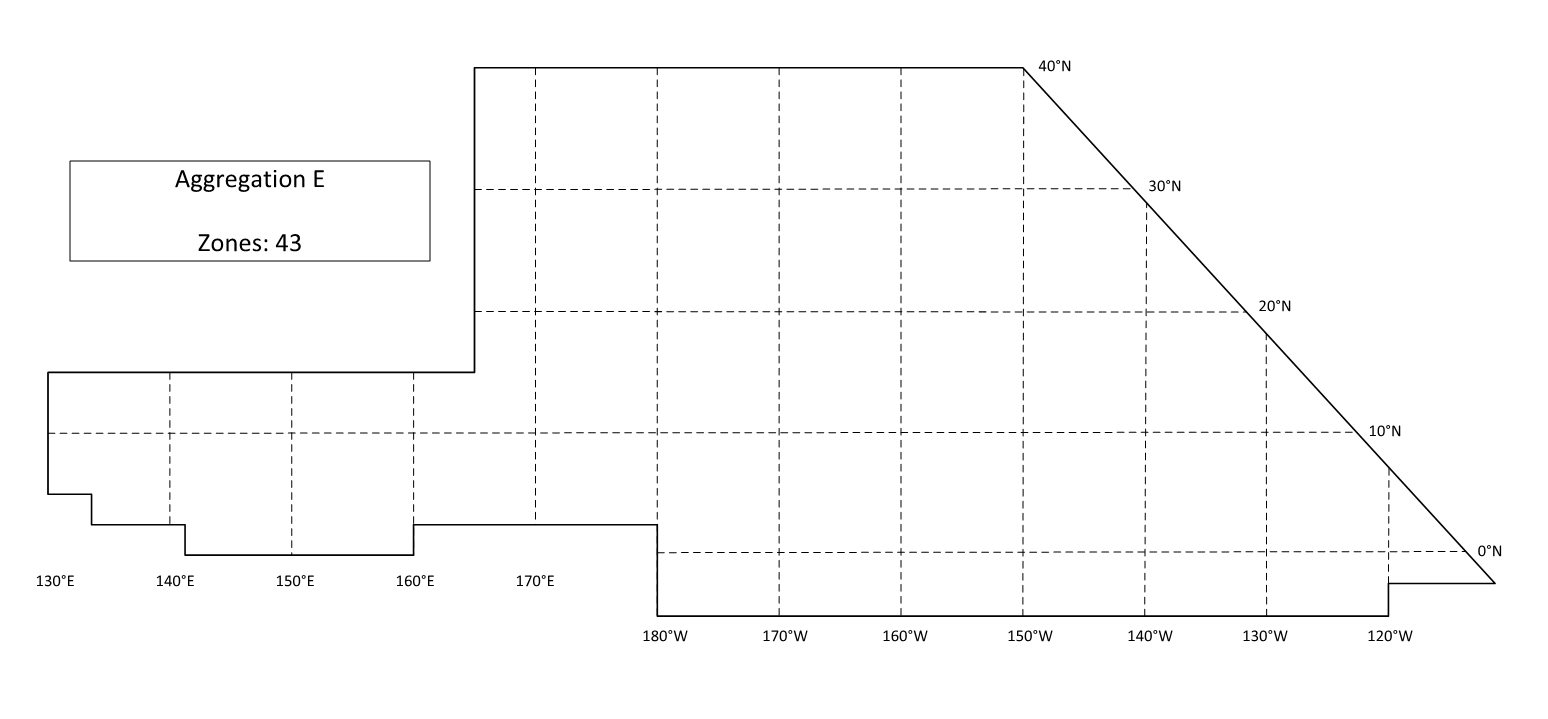}
\caption{Aggregation E (43 Zones)}
\label{AggE}
\end{figure}

\begin{figure}[h!]
\centering
\includegraphics[width=0.85\textwidth]{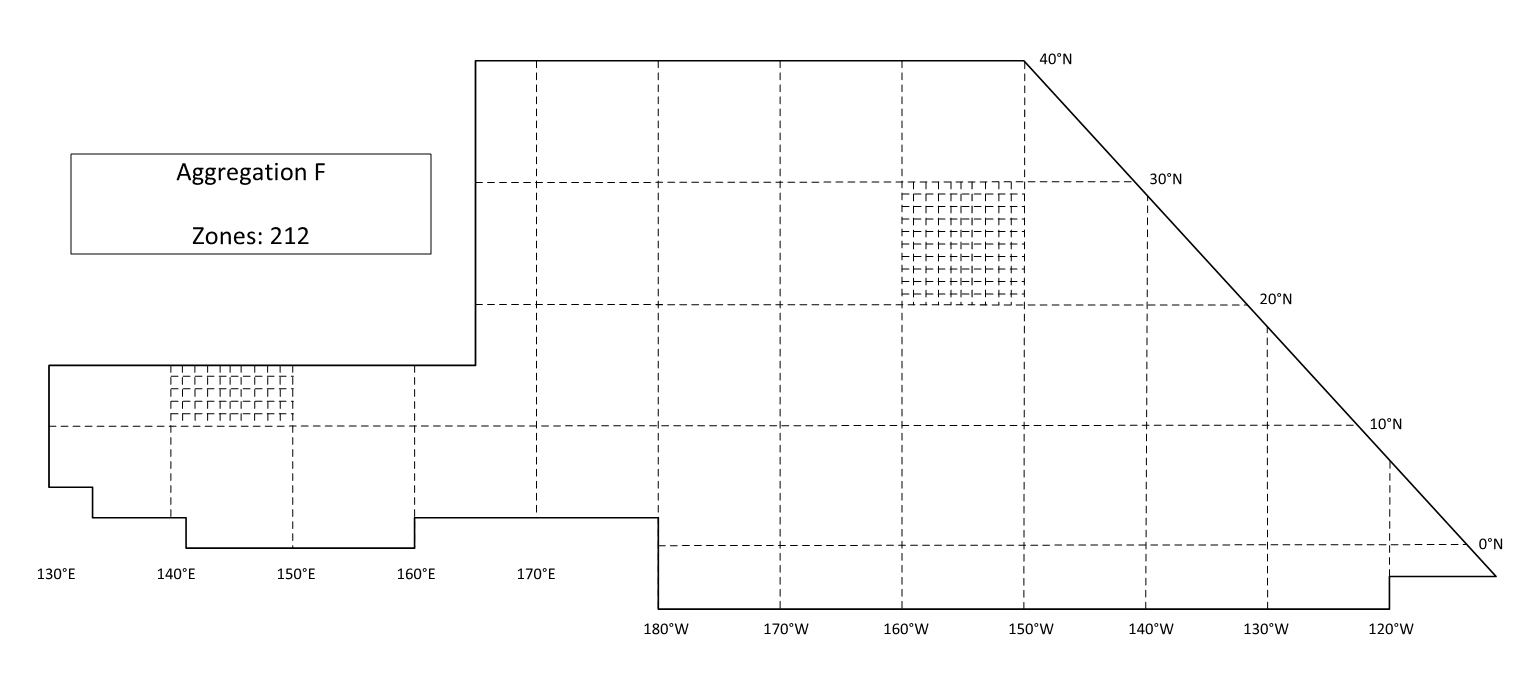}
\caption{Aggregation F (212 Zones)}
\label{AggF}
\end{figure}

 The quadrat aggregation approach was implemented with 6 different quadrat scales to test the impact of the scale effect.  These six grid-based decompositions of the region are labelled Aggregations A - F.  Aggregation A considered the region of study as a singular zone, consolidating all demand points; see Figure \ref{AggA}.  Aggregation B divided the region into two zones along the antimeridiean; see Figure \ref{AggB}.  Aggregations C, D, and E are iterative increases in fidelity, decomposing the region into eight, fifteen, and forty-three zones, respectively; see Figures \ref{AggC}, \ref{AggD}, and \ref{AggE}.  Aggregation F adopts the approach employed by \cite{Akbari} and allows for smaller grid cells in sections of higher demand.  Specifically, the two zones from Aggregation E with the greatest proportion of Guam and Hawaiian Island workloads are further decomposed into $1^o$ x $1^o$ cells; Aggregation F results in 212 zones.  Aggregation F is depicted in Figure \ref{AggF}.

\par
\begin{figure}[h!]
\centering
\includegraphics[width=0.85\textwidth]{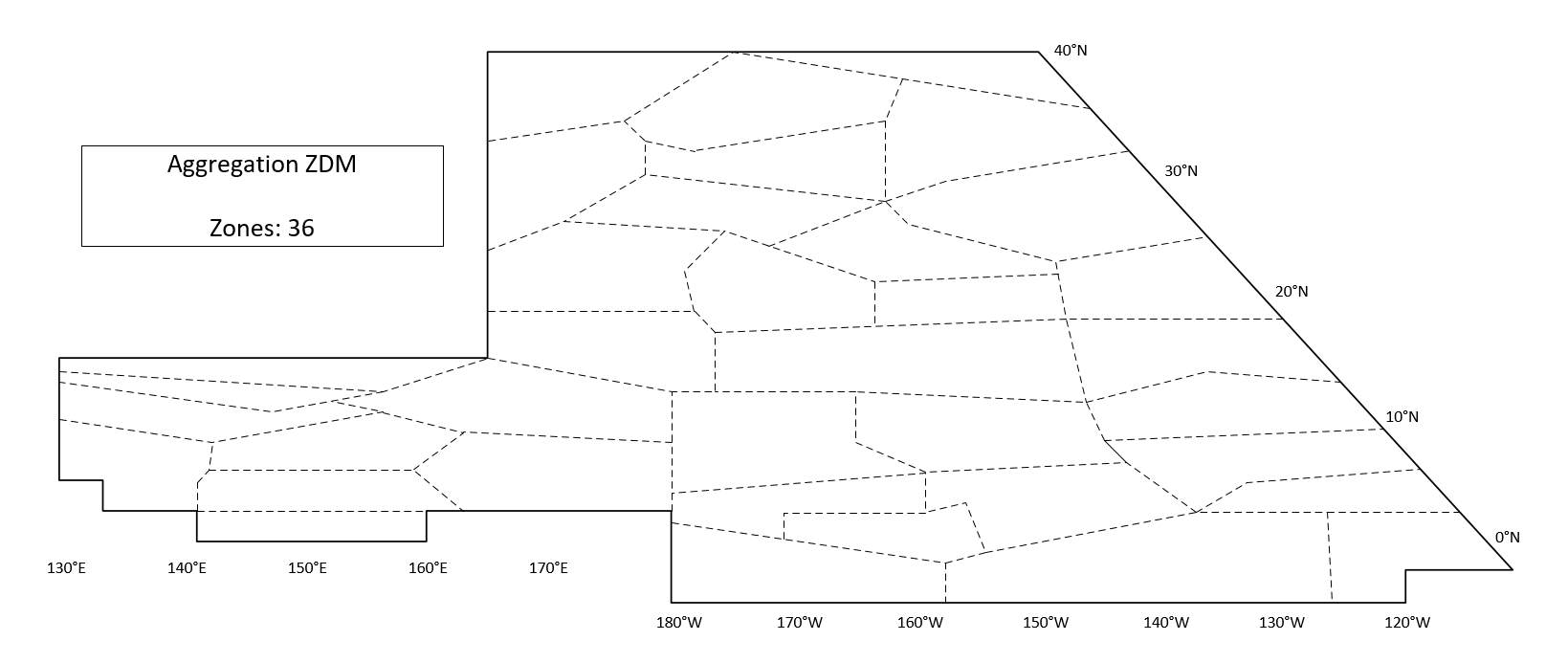}
\caption{Aggregation ZDM (36 Zones)}
\label{ZDM}
\end{figure}

 Aggregation ZDM was constructed utilizing \cite{Razi} general implementation of the zonal distribution model and divided the AOR using a weighted \textit{k}-means clustering algorithm; see Figure \ref{ZDM}.  Razi and Karatas defined the weight of each SAR event using an analytical hierarchy process based upon the level of fatality, material damage, response arduousness, and environmental impact.  Their weighting scheme was not viable for this study based on the available information in MISLE, so this implementation of Razi and Karatas's procedure utilizes \textit{total activities} as a weighting.  The metric of total activities represents the number of resources assigned to a rescue operations, in addition to the instances when a significant change occurred in the course of the rescue operation; this metric of total activities serves as a proxy for the complexity of a SAR event.  Razi and Karatas determine the number of zones to cluster demand points into based upon a \textit{rule of thumb method} proposed by \cite{Kodinariya}.  This method suggests that the number of zones Z is based upon the total number of events K, such that $|Z| \approx \sqrt{|K|/2}$.

\par
\begin{figure}[h!]
\centering
\includegraphics[width=0.85\textwidth]{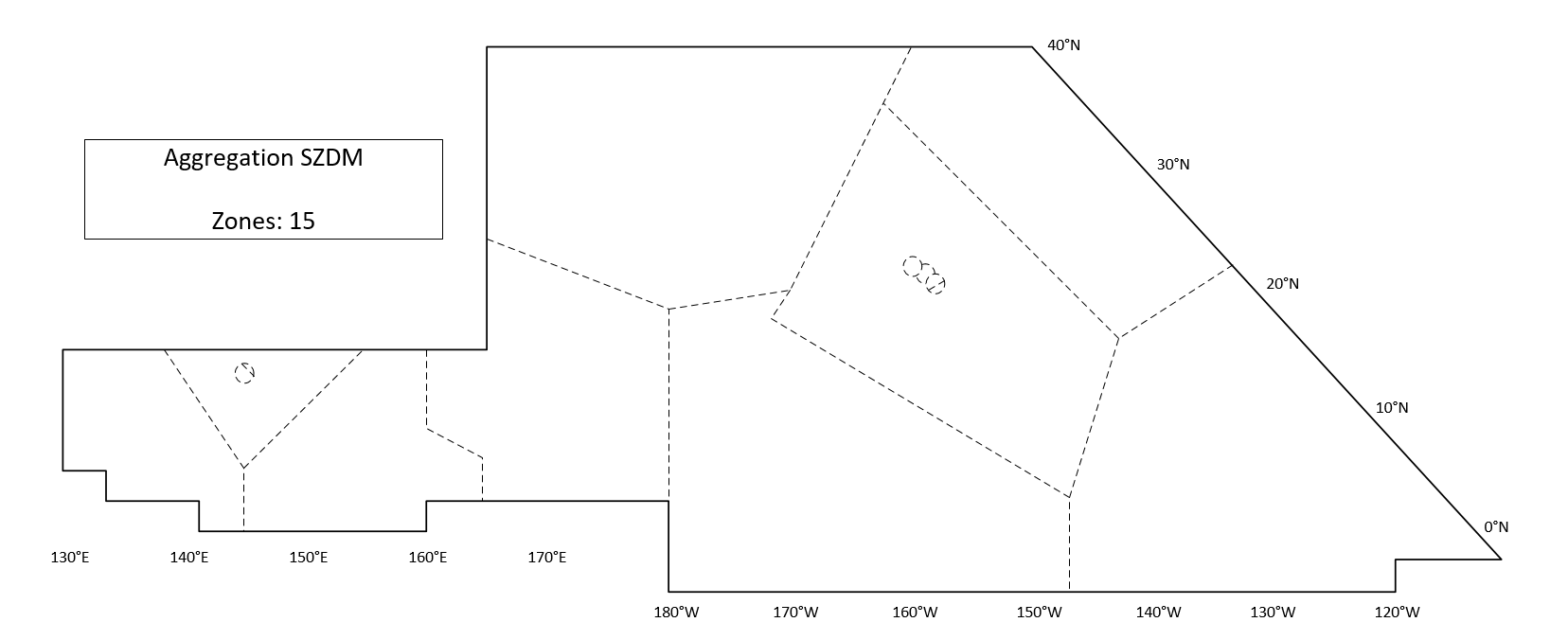}
\caption{Aggregation SZDM (15 Zones)}
\label{SZDM}
\end{figure}

 Aggregation SZDM was developed by implementing the stochastic zonal distribution model approach proposed by \cite{Hornberger}; see Figure \ref{SZDM}.  Hornberger et al. utilized a hierarchical \textit{k}-means clustering algorithm to aggregate demand points into zones.  All demand points are sorted into mutually exclusive groups based upon the unit that coordinated the response and the types of assets utilized in the response.  District 14 is divided into Sector Guam and Sector Honolulu, which split the coverage of the AOR around longitude $160^o$ E.  Current policy dictates that the mission range for USCG boats is 50 nautical miles from the shoreline of an island on which there exists a USCG boat station; District 14 has boat stations located on the islands of Guam, O'ahu, Kaua'i, and Maui.  Hornberger et al. note that a reasonable approximation of asset utilization would be a combination of boats and helicopter aircraft responding to SAR events within the 50 nautical mile boundary of these islands while a combination of cutters and aeroplane aircraft respond to SAR events beyond these boundaries.  Therefore, all demand points where sorted into the following mutually exclusive groups: Guam Boat/Helicopter Events, Guam Cutter/Airplane Events, Hawaii Boat/Helicopter Events, and Hawaii Cutter/Airplane Events.  These groups are further decomposed into clusters based upon the geographic proximity of the data points by employing a \textit{k}-means clustering algorithm.  The number of zones was determined by considering the relationship between the number of zones and the corresponding within-cluster variance.  A plot of this relationship forms an \textit{elbow curve}, whose name is tied to the phenomena that initial groupings account for a greater reduction in variance compared to subsequent groupings; the `elbow' of the curve occurs at the suggested number of zones for the data set.

\subsection{Methods of Comparative Analysis}

\par This study evaluates the effectiveness of various methods of aggregation when conducting spatiotemporal forecasting.  Specifically, we seek to assess the merit of the \cite{Razi} deterministic zonal distribution mode, and the \cite{Hornberger} stochastic zonal distribution model, comparing their effectiveness against traditional quadrat methods of varying fidelity's.  To conduct these comparisons, two metrics are considered: distance-based aggregation error and volume-based aggregation error.

\par The distance-based aggregation error ($d_e$) represents the total distance between where events were modelled as occurring ($\hat{x_j}$) and the actual location of their occurrence ($x_{i, j}$), for each event ($i \in I$) in the zone ($j \in J$).  The anticipated event locations for all zones are weighted centroids for the each zone.  In the quadrat models, the centroids are computed as an average of the latitudes/longitudes, multiplied by the events' corresponding total activities, for all events in the zone.  In the zonal and stochastic zonal distribution models, the clustering algorithm yields a weighted centroid. The distance-based aggregation error metric is:

    \begin{equation} \label{DistMetric}
    d_e = \sum_{i \in I} \sum_{j \in J} |x_{i, j} - x_j|
    \end{equation}
    
\noindent where the Haversine formula, 

    \begin{equation} \label{haversine}
    d = 2r \text{ arcsin} \left( \sqrt{\text{sin}^2\left( \frac{\phi_2 - \phi_1}{2} \right) + \text{cos}(\phi_1)\text{cos}(\phi_2)\text{sin}^2\left( \frac{\theta_2 - \theta_1}{2}\right)} \right)
    \end{equation}

\noindent which, given latitudes $\phi$, and longitudes $\theta$, calculates the great-circle distance between two points, is used to calculate each individual distance. 

\par The weighted distance-based aggregation error ($d_{we}$) is the sum of the differences in distance between where individual assets are modelled as being deployed to ($\hat{x_j}$) and the actual location assets are dispatched to.  The weighting ($w_i$) is the number of assets assigned to the rescue operation.  The difference between $d_e$ and $d_{we}$ is that the former treats individual SAR events as being equal in magnitude, whereas the latter incorporates the number of deployed assets.  As with $d_e$ the individual distances in $d_{we}$ are calculated using the Haversine formula.

\begin{equation} \label{DistMetric}
d_{we} = \sum_{i \in I} \sum_{j \in J} w_i|x_{i, j} - x_j|
\end{equation}

%\par Forecasts of future SAR demand frequently consider the impact that seasonality has on emergency rates.  Autocorrelation analysis and plots comparing SAR activities for each month are used to compare the impact that aggregation has on seasonality measures.

The distance-based aggregation error ($d_e$), and the weighted distance-based aggregation error ($d_{we}$) are both computed for all aggregations A-F, as well as for the ZDM and the SZDM.

\par The volume-based aggregation error ($v_e$) represents the total difference between the predicted level of monthly demand ($\hat{l}_{j, k}$) and the actual level of monthly demand ($l_j$), for each month in the considered time frame ($k \in K$).  The metric is computed as:

\begin{equation} \label{VolMetric}
v_e = \sum_{j \in J} \sum_{k \in K} |l_{j, k} - l_j|
\end{equation}

\par Given that a primary difference between ZDM and SZDM is the integration of stochastic elements in the modeling of the demand, both deterministic and stochastic demand comparisons for volume-based aggregation error are conducted.  For purposes of consistency, all frequency considerations are made on a \textit{per month} basis.

\par Aggregations A-F are compared to the ZDM using a deterministic demand signal.  This requires a singular, static value which represents the typical demand volume for each zone.  Two methods are frequently used to identify these deterministic values: averages and medians.  The average value is a common metric and is familiar to an end-user decision maker, but can be easily skewed by the presence of outliers.  Median values tend to be more stable in the presence of outliers and thus more representative of the typical demand volume.  As such, median values are implemented as the metric for deterministic demand volume in this study.

\par The stochastic modelling approach utilized in SZDM considers the inherent uncertainty present in SAR events by fitting probability distributions to demand volumes in each zone.  As noted by \cite{Afshartous} and \cite{Akbari}, SAR events can often be viewed as Poisson processes.  In particular, \cite{Hornberger} found the emergence of SAR events in District 14's AOR could be modelled using poisson and gamma-poisson distributions.  This study implements stochastic demand modeling in SZDM, and compares this to aggregations C and D to compare the impact of aggregation method on the simulation of future SAR demand. (Aggregations A and B were deemed too trivial to be of real interest, and stochastic models of Aggregations E and F proved intractable on the authors' hardware.)

\par   A modification of the volume-based aggregation error, $v_e = \sum_{j \in J} \sum_{k \in K} ( l_{j, k} - l_j)$, is also considered providing a distinction between over- and under-forecasting events.  Stochastic models are compared graphically, plotting the simulated output for each month of the 24-month test period against the actual demand volume observed.

\section{Analysis}

\subsection{Distance-Based Aggregation Error}

\par The distances, in nautical miles, between the aggregated demand point and the subsequent demand nodes during 2016 - 2017 are shown in Table \ref{DeterDemandDist}.  The resulting distance-based aggregation error for the quadrat models reflect the law of diminishing returns, as described by \cite{Francis2004}.  The first division of the region of study, from Aggregation A to Aggregation B, results in an 82.3\% reduction to the locational aggregation error.  This error was continuously diminished with additional divisions.  These results support the trend of location error generally reducing with additional zones.

    \begin{table}[h!]
    \centering
    \small
    \caption{Distance-Based Aggregation Error}
    \label{DeterDemandDist}    
    \begin{tabular}{| c | c | c | c |}
    \hline
        \textbf{Aggregation} & \textbf{Number of Zones} & $d_e$ & $d_{we}$ \\
        \hline
        A & 1 & 1,471,479 & 2,195,276 \\
        B & 2 & 251,042.3 & 312,118.6 \\
        C & 8 & 171,531.3 & 225,615 \\
        D & 15 & 158,119.1 & 208,812.7 \\
        E & 43 & 86,745.88 & 119,741.5 \\
        F & 212 & 51,553.33 & 66,668.67 \\
        \textit{ZDM} & \textit{36} & \textit{80,165.06} & \textit{92,669.37} \\
        \textit{SZDM} & \textit{15} & \textit{92,067.72} & \textit{97,425.77} \\
    \hline
    \end{tabular}
    \end{table} 

\par Aggregations ZDM and SZDM perform very well compared to the quadrat models.  The zonal distribution model has a lower associated location error than Aggregation E, despite only having 36 zones compared to Aggregation E's 43 zones.  This runs counter to the general claim that more zones always improves the accuracy of the location model, suggesting instead that deliberate steps can be implemented to aggregate spatial demand points in fewer clusters while still achieve competitively low levels of location error.  The stochastic zonal distribution model's results support this observation, achieving a 41.7\% reduction in distance-based aggregation error compared to Aggregation D despite using the same number of zones.

\par Similar trends are observed when the attention is shifted from the error in SAR event distances to the error in resource dispatch distances.  There is a steady improvement in accuracy as the number of zones is increased, with the exception of Aggregations ZDM and SZDM.  Additionally, the differences between $d_e$ and $d_{we}$ are notably larger for the quadrat models compared to Aggregations ZDM and SZDM; the stochastic zonal distribution model had the smallest increase in location error when weighting by the number of resources dispatched.  These observations suggest that deliberate zoning of demand point can enhance the robustness of aggregate zones to weighted events, particularly when the zones are developed with consideration to both geographic proximity and the operational characteristics that are tied to the event weights.

\subsection{Deterministic Volume-Based Aggregation Error}

\par The total error in volume based upon the median monthly demand for each zone compared to the actual demand volumes as depicted in Table \ref{DeterDemandVol}.  The phenomena described by \cite{Francis2008} and \cite{Dark2007} is observed; there is a general increase in total volume-based aggregation error as the number of zones increases.

    \begin{table}[h!]
    \centering
    \small
    \caption{Volume-Based Aggregation Error for Deterministic Demand Modeling}
    \label{DeterDemandVol}    
    \begin{tabular}{| c | c | c |}
    \hline
        \textbf{Aggregation} & \textbf{Number of Zones} & $v_e$ \\
        \hline
        A & 1 & 139 \\
        B & 2 & 189 \\
        C & 8 & 288 \\
        D & 15 & 306 \\
        E & 44 & 372 \\
        F & 212 & 584 \\
        \textit{ZDM} & \textit{36} & \textit{458} \\
    \hline
    \end{tabular}
    \end{table} 

\par
\begin{figure}[h!]
\centering
\includegraphics[width=1\textwidth]{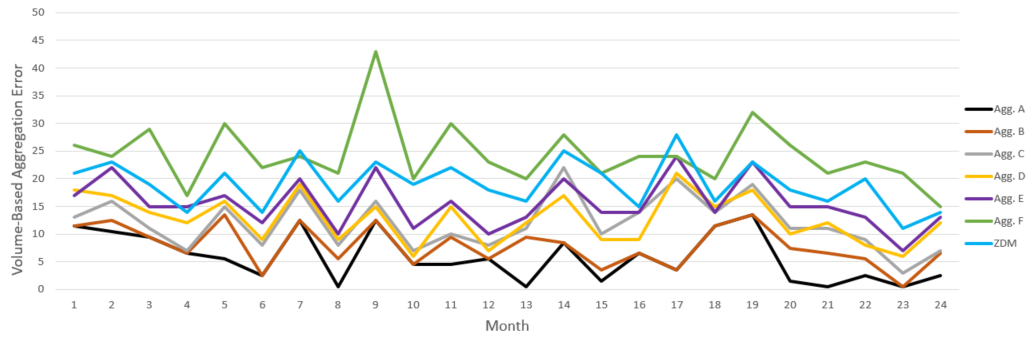}
\caption{Comparison of the Total Volume-Based Aggregation Error for Deterministic Demand Modeling}
\label{Total_Vol}
\end{figure}

 Interestingly, implementing the zonal distribution model corresponds to a large volume-based aggregation area, second only to Aggregation F; see Figure \ref{Total_Vol}.  This suggests deliberate clustering based on geographic proximity does not correspond to improvements in deterministic demand volume modeling.

\par
\begin{figure}[h!]
\centering
\includegraphics[width=1\textwidth]{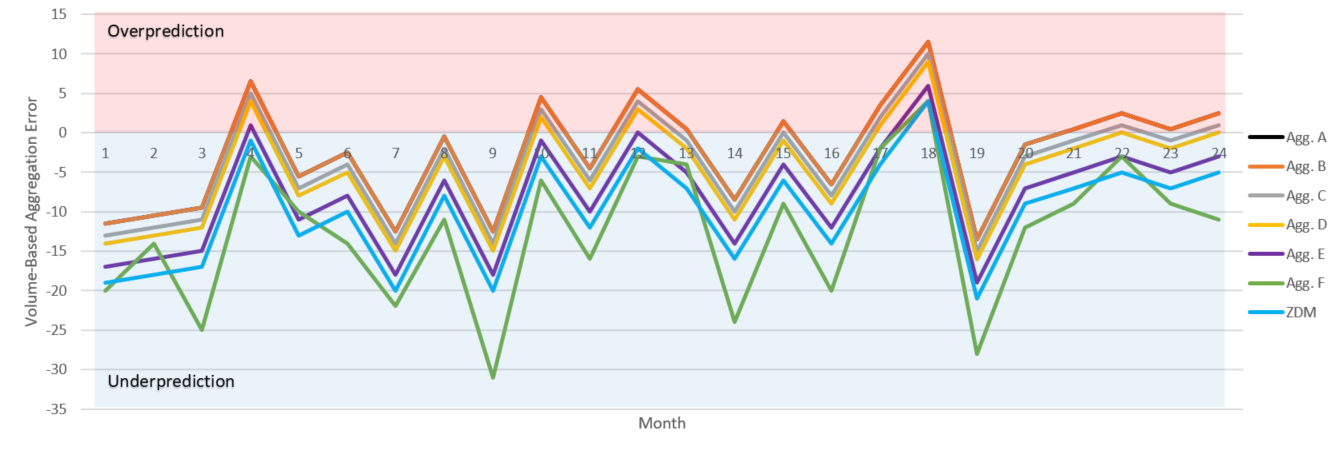}
\caption{Comparison of Over- and Under-predictions fo Deterministic Demand Modeling}
\label{OverUnder}
\end{figure}

 Additional analysis compared the tendency for different aggregation models to overpredict versus underpredict demand volume.  A plot of this analysis is shown in Figure \ref{OverUnder}, colorcoding the region of overprediction as red and underprediction as blue.  For each month, Aggregation A and B perform equally well; the lines overlap in the plot.  With the exception of Aggregation F, all methods adhere to similar trends in spikes and drops throughout the test timeframe.  The general trend is for models to underpredict more consistently as they incorporate more aggregated zones.  The exception to this trend is the zonal distribution model, which continues to have greater volume-based aggregation error compared to Aggregation E.

\subsection{Stochastic Volume-Based Aggregation Error}

\par
\begin{figure}[h!]
\centering
\includegraphics[width=1\textwidth]{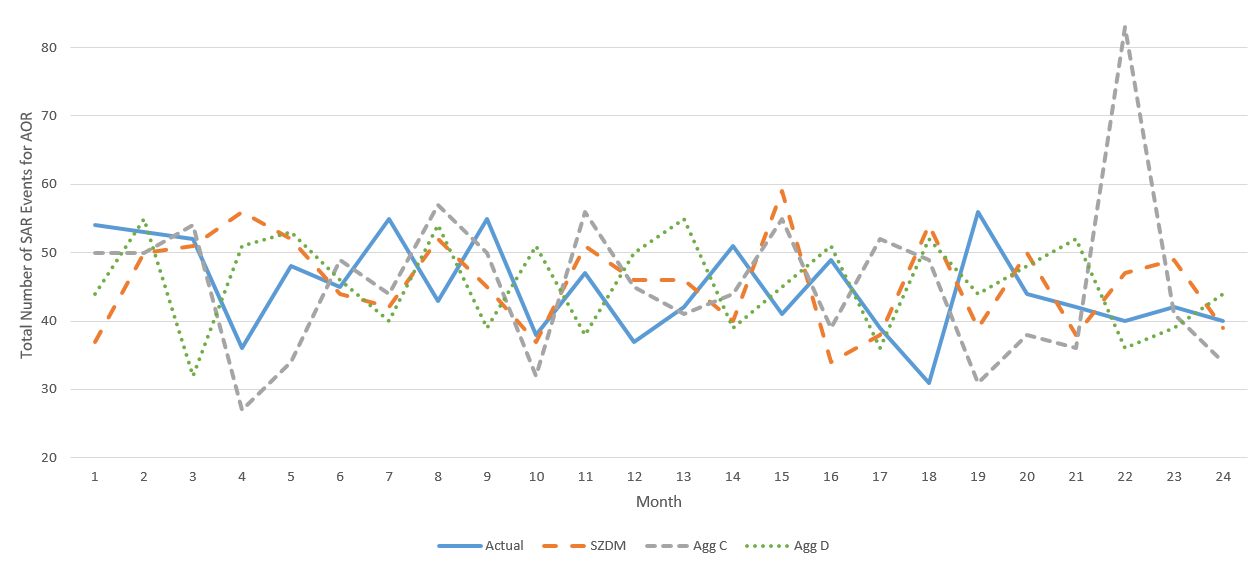}
\caption{Comparison of Stochastic Demand Models and Observed Demand Levels}
\label{Stoch_Vol}
\end{figure}

 A comparison of stochastic demand models was used probability distributions fit to each zone in Aggregations C, D, and SZDM.  The results from these simulations are compared to the actual observed demand levels for the two-year test period; see Figure \ref{Stoch_Vol}.  Note that since the demand distributions were observed to be relatively stationary at large, each month's simulated volume from each model is determined by random draws from static probability distributions assigned to each zone (i.e., poisson and gamma-poisson distributions).

\par Since the results from Figure \ref{Stoch_Vol} are randomly generated, the emphasis is less on the specific results from month-to-month and more on whether overall trend appears similar to the observed trend.  This analysis shows similar trends for the three stochastic demand models, suggesting that they all could be used to effectively simulate the stochastic demand of the AOR.  Aggregation C does make a notable spike in simulated SAR activity at the end of the test period, caused by the coincidence of multiple zones within the model simulating larger-than-normal demand volume.  This phenomena was investigated further.

\par While the observed demand volume fluctuates from month-to-month, it stays within the bounds of 30 and 60 events per month.  Using these levels as thresholds, a monte carlo simulation of 10,000 2-year models was constructed.  For each of the 240,000 simulated months, Table \ref{ExtMonth} shows the number that were beyond the thresholds of 30 and 60 events per month.  All models appear relatively stable compared to these bounds; Aggregation C, with the greatest number of `extreme months', only had approximately 4.6\% of the 240,000 months classified as `extreme'.  The stochastic zonal distribution model appeared to be the most stable of the three considered models, having the fewest months classified as `extreme' on either side of the bound.  These findings suggests that while extreme months are not likely to be a significant occurrence in a simulation of SAR demand, the stochastic zonal distribution model minimizes the likelihood this will occur.

\begin{table}[h!]
    \centering
    \small
    \caption{Comparison of Extreme Months over 10,000 2-Year Simulations}
    \label{ExtMonth}    
    \begin{tabular}{| c | c | c |}
    \hline
        \textbf{Aggregation} & \textbf{Below 30 Events} & \textbf{Above 60 Events} \\
        \hline
        C & 6175 & 5056 \\
        D & 5402 & 4660 \\
        \textit{SZDM} & 4727 & 3854 \\
    \hline
    \end{tabular}
\end{table} 
    
\section{Conclusion}

\par The method used to aggregate spatiotemporal demands affects the outcome of location models built using the aggregated data, thus an understanding of the impacts of aggregation methods is fundamental. We have presented a framework for comparison of both static and stochastic spatiotemporal aggregation models, utilizing both a distance based aggregation error metric, an event magnitude weighted distance based aggregation error metric, and a volume based aggregation error metric.  We further applied this framework to test six quadrat aggregation models of varying fidelity's, and two zonal based models, using historical search and rescue data from a massive scale region possessing highly variable demands. As expected aggregations with greater fidelity tend to reduce the distance-based aggregation error.  In addition implementation of a deliberate zoning approach (e.g., ZDM and SDZM) further reduce this error while utilizing fewer zones. However, higher fidelity aggregations with increased number of zones has a detrimental effect on the modelling of demand volumes. Finally, stochastic representations of SAR demand appears to be effective at simulating actual SAR demand.

\par Based on the results of our aggregation analysis we propose the following as potential exploratory  efforts.  Zonal techniques based on hierarchies and clustering techniques seem very promising, additional research on the impacts of clustering techniques could be fruitful. Additionally combining these zonal techniques, with their associated reduced location errors, with a lower fidelity aggregation model to project region level demands may be useful. Finally, a study examining possible nonlinear dynamic effects on the resulting output of location models as a result of changes in aggregation method may be informative.  

%\section*{Disclaimer} The views expressed in this article are those of the authors and do not reflect the official policy or position of the United States Air Force, the Department of Defense, or the United States Government.

%\section*{Acknowledgements}
%Funding: This work was supported by the the United States Coast Guard Research and Development Center.

\bibliography{mybibfile}

\begin{thebibliography}{35}
\expandafter\ifx\csname natexlab\endcsname\relax\def\natexlab#1{#1}\fi
\providecommand{\url}[1]{\texttt{#1}}
\providecommand{\href}[2]{#2}
\providecommand{\path}[1]{#1}
\providecommand{\DOIprefix}{doi:}
\providecommand{\ArXivprefix}{arXiv:}
\providecommand{\URLprefix}{URL: }
\providecommand{\Pubmedprefix}{pmid:}
\providecommand{\doi}[1]{\href{http://dx.doi.org/#1}{\path{#1}}}
\providecommand{\Pubmed}[1]{\href{pmid:#1}{\path{#1}}}
\providecommand{\bibinfo}[2]{#2}
\ifx\xfnm\undefined \def\xfnm[#1]{\unskip,\space#1}\fi
%Type = Article
\bibitem[{Afshartous et~al.(2009)Afshartous, Guan and Mehrotra}]{Afshartous}
\bibinfo{author}{Afshartous\xfnm[ D.]}, \bibinfo{author}{Guan\xfnm[ Y.]},
  \bibinfo{author}{Mehrotra\xfnm[ A.]}.
\newblock \bibinfo{title}{Us coast guard air station location with respect to
  distress calls: a spatial statistics and optimization based methodology}.
\newblock \bibinfo{journal}{European Journal of Operational Research}
  \bibinfo{year}{2009};\bibinfo{volume}{196}:\bibinfo{pages}{1086--1096}.
\newblock \DOIprefix\doi{10.1016/j.ejor.2008.04.010}.
%Type = Article
\bibitem[{Ai et~al.(2015)Ai, Lu and Zhang}]{Ai}
\bibinfo{author}{Ai\xfnm[ Y.]}, \bibinfo{author}{Lu\xfnm[ J.]},
  \bibinfo{author}{Zhang\xfnm[ L.]}.
\newblock \bibinfo{title}{The optimization model for the location of maritime
  emergency supplies reserve bases and the configuration of salvage vessels}.
\newblock \bibinfo{journal}{Transportation Research Part E: Logistics and
  Transportation Review}
  \bibinfo{year}{2015};\bibinfo{volume}{83}:\bibinfo{pages}{170--188}.
\newblock \DOIprefix\doi{10.1016/j.tre.2015.09.006}.
%Type = Article
\bibitem[{Akbari et~al.(2018{\natexlab{a}})Akbari, Eiselt and
  Pelot}]{AkbariINFOR}
\bibinfo{author}{Akbari\xfnm[ A.]}, \bibinfo{author}{Eiselt\xfnm[ H.]},
  \bibinfo{author}{Pelot\xfnm[ R.]}.
\newblock \bibinfo{title}{A maritime search and rescue location analysis
  considering multiple criteria, with simulated demand}.
\newblock \bibinfo{journal}{INFOR: Information Systems and Operations Research}
  \bibinfo{year}{2018}{\natexlab{a}};\bibinfo{volume}{56:1}:\bibinfo{pages}{92--114}.
%Type = Article
\bibitem[{Akbari et~al.(2018{\natexlab{b}})Akbari, Pelot and Eiselt}]{Akbari}
\bibinfo{author}{Akbari\xfnm[ A.]}, \bibinfo{author}{Pelot\xfnm[ R.]},
  \bibinfo{author}{Eiselt\xfnm[ H.]}.
\newblock \bibinfo{title}{A modular capacitated multi-objective model for
  locating maritime search and rescue vessels}.
\newblock \bibinfo{journal}{Annals of Operations Research}
  \bibinfo{year}{2018}{\natexlab{b}};\bibinfo{volume}{267}:\bibinfo{pages}{3--28}.
\newblock \DOIprefix\doi{10.1007/s10479-017-2593-1}.
%Type = Article
\bibitem[{Anselin et~al.(2000)Anselin, Cohen, Cook, Gorr and Tita}]{Anselin}
\bibinfo{author}{Anselin\xfnm[ L.]}, \bibinfo{author}{Cohen\xfnm[ J.]},
  \bibinfo{author}{Cook\xfnm[ D.]}, \bibinfo{author}{Gorr\xfnm[ W.]},
  \bibinfo{author}{Tita\xfnm[ G.]}.
\newblock \bibinfo{title}{Spatial analysis of crime}.
\newblock \bibinfo{journal}{Measurement and Analysis of Crime and Justice}
  \bibinfo{year}{2000};\bibinfo{volume}{vol 4}:\bibinfo{pages}{213--262}.
%Type = Article
\bibitem[{Araz et~al.(2007)Araz, Selim and Ozkarahan}]{araz2007fuzzy}
\bibinfo{author}{Araz\xfnm[ C.]}, \bibinfo{author}{Selim\xfnm[ H.]},
  \bibinfo{author}{Ozkarahan\xfnm[ I.]}.
\newblock \bibinfo{title}{A fuzzy multi-objective covering-based vehicle
  location model for emergency services}.
\newblock \bibinfo{journal}{Computers \& Operations Research}
  \bibinfo{year}{2007};\bibinfo{volume}{34}(\bibinfo{number}{3}):\bibinfo{pages}{705--726}.
%Type = Article
\bibitem[{Armstrong and Cook(1979)}]{Armstrong}
\bibinfo{author}{Armstrong\xfnm[ R.]}, \bibinfo{author}{Cook\xfnm[ W.]}.
\newblock \bibinfo{title}{Goal programming models for assigning search and
  rescue aircraft to bases}.
\newblock \bibinfo{journal}{The Journal of the Operational Research Society}
  \bibinfo{year}{1979};\bibinfo{volume}{30}:\bibinfo{pages}{555--561}.
%Type = Article
\bibitem[{Azofra et~al.(2007)Azofra, Perez-Labajos, Blanco and
  Achutegui}]{Azofra}
\bibinfo{author}{Azofra\xfnm[ M.]}, \bibinfo{author}{Perez-Labajos\xfnm[ C.]},
  \bibinfo{author}{Blanco\xfnm[ B.]}, \bibinfo{author}{Achutegui\xfnm[ J.]}.
\newblock \bibinfo{title}{Optimum placement of sea rescue resources}.
\newblock \bibinfo{journal}{Safety Science}
  \bibinfo{year}{2007};\bibinfo{volume}{45}:\bibinfo{pages}{941--951}.
\newblock \DOIprefix\doi{10.1016/j.ssci.2006.09.002}.
%Type = Article
\bibitem[{Chainey and Dando(2005)}]{Chainey}
\bibinfo{author}{Chainey\xfnm[ S.]}, \bibinfo{author}{Dando\xfnm[ J.]}.
\newblock \bibinfo{title}{Chapter 2. methods and techniques for understanding
  crime hot spots}.
\newblock \bibinfo{journal}{Mapping Crime: Understanding Hot Spots}
  \bibinfo{year}{2005};:\bibinfo{pages}{15--34}.
%Type = Article
\bibitem[{Current and Schilling(1987)}]{Current1987}
\bibinfo{author}{Current\xfnm[ J.]}, \bibinfo{author}{Schilling\xfnm[ D.]}.
\newblock \bibinfo{title}{Elimination of source a and b errors in p-median
  location problems}.
\newblock \bibinfo{journal}{Geographical Analysis}
  \bibinfo{year}{1987};\bibinfo{volume}{19}.
%Type = Article
\bibitem[{Current and Schilling(1990)}]{Current1990}
\bibinfo{author}{Current\xfnm[ J.]}, \bibinfo{author}{Schilling\xfnm[ D.]}.
\newblock \bibinfo{title}{Analysis of errors due to demand data aggregation in
  the set covering and maximal covering location problems}.
\newblock \bibinfo{journal}{Geographical Analysis}
  \bibinfo{year}{1990};\bibinfo{volume}{22}:\bibinfo{pages}{116--126}.
%Type = Article
\bibitem[{Curtis and MacPherson(1996)}]{Curtis1996}
\bibinfo{author}{Curtis\xfnm[ A.]}, \bibinfo{author}{MacPherson\xfnm[ A.]}.
\newblock \bibinfo{title}{The zone definition problem in survey research: an
  empirical example from new york state}.
\newblock \bibinfo{journal}{Professional Geographer}
  \bibinfo{year}{1996};\bibinfo{volume}{48}:\bibinfo{pages}{310--323}.
%Type = Article
\bibitem[{Dark and Bram(2007)}]{Dark2007}
\bibinfo{author}{Dark\xfnm[ S.]}, \bibinfo{author}{Bram\xfnm[ D.]}.
\newblock \bibinfo{title}{The modifiable areal unit problem (maup) in physical
  geography}.
\newblock \bibinfo{journal}{Progress in Physical Geography}
  \bibinfo{year}{2007};\bibinfo{volume}{31}:\bibinfo{pages}{471--479}.
%Type = Article
\bibitem[{Erdemir et~al.(2008)Erdemir, Batta, Spielman, Rogerson, Blatt and
  Flanigan}]{Erdemir}
\bibinfo{author}{Erdemir\xfnm[ E.]}, \bibinfo{author}{Batta\xfnm[ R.]},
  \bibinfo{author}{Spielman\xfnm[ S.]}, \bibinfo{author}{Rogerson\xfnm[ P.]},
  \bibinfo{author}{Blatt\xfnm[ A.]}, \bibinfo{author}{Flanigan\xfnm[ M.]}.
\newblock \bibinfo{title}{Optimization of aeromedical base locations in new
  mexico using a model that considers crash nodes and paths}.
\newblock \bibinfo{journal}{Accident Analysis and Prevention}
  \bibinfo{year}{2008};\bibinfo{volume}{40}:\bibinfo{pages}{1105--1114}.
%Type = Article
\bibitem[{Fotheringham et~al.(1995)Fotheringham, Densham and
  Curtis}]{Fotheringham1995}
\bibinfo{author}{Fotheringham\xfnm[ A.]}, \bibinfo{author}{Densham\xfnm[ P.]},
  \bibinfo{author}{Curtis\xfnm[ A.]}.
\newblock \bibinfo{title}{The zone definition problem in location-allocation
  modeling}.
\newblock \bibinfo{journal}{Geographical Analysis}
  \bibinfo{year}{1995};\bibinfo{volume}{27}:\bibinfo{pages}{60--77}.
%Type = Article
\bibitem[{Francis and Lowe(1992)}]{Francis1992}
\bibinfo{author}{Francis\xfnm[ R.]}, \bibinfo{author}{Lowe\xfnm[ T.]}.
\newblock \bibinfo{title}{On worst-case aggregation analysis for network
  location problems}.
\newblock \bibinfo{journal}{Annals of Operations Research}
  \bibinfo{year}{1992};\bibinfo{volume}{40}:\bibinfo{pages}{229--246}.
%Type = Article
\bibitem[{Francis and Lowe(2014)}]{Francis2014}
\bibinfo{author}{Francis\xfnm[ R.]}, \bibinfo{author}{Lowe\xfnm[ T.]}.
\newblock \bibinfo{title}{Comparative error bound theory for three location
  models: continuous demand versus discrete demand}.
\newblock \bibinfo{journal}{TOP}
  \bibinfo{year}{2014};\bibinfo{volume}{1}:\bibinfo{pages}{144--169}.
%Type = Article
\bibitem[{Francis et~al.(2008)Francis, Lowe, Rayco and Tamir}]{Francis2008}
\bibinfo{author}{Francis\xfnm[ R.]}, \bibinfo{author}{Lowe\xfnm[ T.]},
  \bibinfo{author}{Rayco\xfnm[ M.]}, \bibinfo{author}{Tamir\xfnm[ A.]}.
\newblock \bibinfo{title}{Aggregation error for location models: survey and
  analysis}.
\newblock \bibinfo{journal}{Annals of Operations Research}
  \bibinfo{year}{2008};\bibinfo{volume}{167}:\bibinfo{pages}{171--208}.
%Type = Article
\bibitem[{Francis et~al.(2004{\natexlab{a}})Francis, Lowe and
  Tamir}]{Francis2004b}
\bibinfo{author}{Francis\xfnm[ R.]}, \bibinfo{author}{Lowe\xfnm[ T.]},
  \bibinfo{author}{Tamir\xfnm[ A.]}.
\newblock \bibinfo{title}{Demand point aggregation analysis for a class of
  constrained location models: a penalty function approach}.
\newblock \bibinfo{journal}{IIE Transactions}
  \bibinfo{year}{2004}{\natexlab{a}};\bibinfo{volume}{36}:\bibinfo{pages}{601--609}.
%Type = Article
\bibitem[{Francis et~al.(2004{\natexlab{b}})Francis, Lowe, Tamir and
  Emir-Farinas}]{Francis2004}
\bibinfo{author}{Francis\xfnm[ R.]}, \bibinfo{author}{Lowe\xfnm[ T.]},
  \bibinfo{author}{Tamir\xfnm[ A.]}, \bibinfo{author}{Emir-Farinas\xfnm[ H.]}.
\newblock \bibinfo{title}{Aggregation decomposition and aggregation guidelines
  for a class of minimax and covering location models}.
\newblock \bibinfo{journal}{Geographical Analysis}
  \bibinfo{year}{2004}{\natexlab{b}};\bibinfo{volume}{36(4)}:\bibinfo{pages}{332--349}.
\newblock \DOIprefix\doi{· 10.1111/j.1538-4632.2004.tb01140.x}.
%Type = Article
\bibitem[{Gehlke and Biehl(1934)}]{Gehlke1934}
\bibinfo{author}{Gehlke\xfnm[ C.]}, \bibinfo{author}{Biehl\xfnm[ K.]}.
\newblock \bibinfo{title}{Certain effects of grouping upon the size and
  correlation coefficient in census tract material}.
\newblock \bibinfo{journal}{Journal of the American Statistical Association}
  \bibinfo{year}{1934};\bibinfo{volume}{29}.
%Type = Article
\bibitem[{Hillsman and Rhoda(1978)}]{Hillsman1978}
\bibinfo{author}{Hillsman\xfnm[ E.]}, \bibinfo{author}{Rhoda\xfnm[ R.]}.
\newblock \bibinfo{title}{Errors in measuring distances from populations to
  service centers}.
\newblock \bibinfo{journal}{Annals of Regional Science}
  \bibinfo{year}{1978};\bibinfo{volume}{1}:\bibinfo{pages}{74--88}.
%Type = Article
\bibitem[{Hornberger et~al.(2019)Hornberger, Cox and Lunday}]{Hornberger}
\bibinfo{author}{Hornberger\xfnm[ Z.]}, \bibinfo{author}{Cox\xfnm[ B.]},
  \bibinfo{author}{Lunday\xfnm[ B.]}.
\newblock \bibinfo{title}{Optimal heterogeneous asset location modeling for
  expected spatiotemporal search and rescue demands using historic event data}.
\newblock \bibinfo{journal}{Manuscript submitted for publication}
  \bibinfo{year}{2019};\URLprefix \url{https://arxiv.org/abs/1908.08970}.
%Type = Article
\bibitem[{Jelinsky and Wu(1996)}]{Jelinsky&Wu}
\bibinfo{author}{Jelinsky\xfnm[ D.]}, \bibinfo{author}{Wu\xfnm[ J.]}.
\newblock \bibinfo{title}{The modifiable areal unit problem and implications
  for landscape ecology}.
\newblock \bibinfo{journal}{Landscape Ecology}
  \bibinfo{year}{1996};\bibinfo{volume}{11}:\bibinfo{pages}{129--140}.
%Type = Article
\bibitem[{Karatas et~al.(2017)Karatas, Razi and Gunal}]{Karatas}
\bibinfo{author}{Karatas\xfnm[ M.]}, \bibinfo{author}{Razi\xfnm[ N.]},
  \bibinfo{author}{Gunal\xfnm[ M.]}.
\newblock \bibinfo{title}{An ilp and simulation model to optimize search and
  rescue helicopter operations}.
\newblock \bibinfo{journal}{Journal of the Operational Research Society}
  \bibinfo{year}{2017};\bibinfo{volume}{68}:\bibinfo{pages}{1335--1351}.
\newblock \DOIprefix\doi{10.1057/s41274-016-0154-7}.
%Type = Article
\bibitem[{Kodinariya and Makwana(2013)}]{Kodinariya}
\bibinfo{author}{Kodinariya\xfnm[ T.]}, \bibinfo{author}{Makwana\xfnm[ P.]}.
\newblock \bibinfo{title}{Review on determining number of cluster in k-means
  clustering}.
\newblock \bibinfo{journal}{International Journal of Advance Research in
  Computer Science and Management Studies}
  \bibinfo{year}{2013};\bibinfo{volume}{1(6)}:\bibinfo{pages}{90--95}.
%Type = Article
\bibitem[{Li and Szeto(2019)}]{li2019taxi}
\bibinfo{author}{Li\xfnm[ B.]}, \bibinfo{author}{Szeto\xfnm[ W.]}.
\newblock \bibinfo{title}{Taxi service area design: Formulation and analysis}.
\newblock \bibinfo{journal}{Transportation Research Part E: Logistics and
  Transportation Review}
  \bibinfo{year}{2019};\bibinfo{volume}{125}:\bibinfo{pages}{308--333}.
%Type = Book
\bibitem[{Openshaw(1984)}]{Openshaw}
\bibinfo{author}{Openshaw\xfnm[ S.]}.
\newblock \bibinfo{title}{The modifiable areal unit problem}.
\newblock \bibinfo{address}{Norwich}: \bibinfo{publisher}{Geo Books},
  \bibinfo{year}{1984}.
%Type = Article
\bibitem[{Papadimitriou(1981)}]{Papadimitriou1981}
\bibinfo{author}{Papadimitriou\xfnm[ C.]}.
\newblock \bibinfo{title}{Worst case and probabilistic analysis of a geometric
  location problem}.
\newblock \bibinfo{journal}{SIAM Journal on Computing}
  \bibinfo{year}{1981};\bibinfo{volume}{1}.
%Type = Article
\bibitem[{Qi and Shen(2010)}]{Qi2010}
\bibinfo{author}{Qi\xfnm[ L.]}, \bibinfo{author}{Shen\xfnm[ Z.]}.
\newblock \bibinfo{title}{Worst-case analysis of demand point aggregation for
  the euclidean p-median problem}.
\newblock \bibinfo{journal}{European Journal of Operational Research}
  \bibinfo{year}{2010};\bibinfo{volume}{202}:\bibinfo{pages}{434--443}.
%Type = Article
\bibitem[{Rajendran and Zack(2019)}]{rajendran2019insights}
\bibinfo{author}{Rajendran\xfnm[ S.]}, \bibinfo{author}{Zack\xfnm[ J.]}.
\newblock \bibinfo{title}{Insights on strategic air taxi network infrastructure
  locations using an iterative constrained clustering approach}.
\newblock \bibinfo{journal}{Transportation Research Part E: Logistics and
  Transportation Review}
  \bibinfo{year}{2019};\bibinfo{volume}{128}:\bibinfo{pages}{470--505}.
%Type = Article
\bibitem[{Razi and Karatas(2016)}]{Razi}
\bibinfo{author}{Razi\xfnm[ N.]}, \bibinfo{author}{Karatas\xfnm[ M.]}.
\newblock \bibinfo{title}{A multi-objective model for locating search and
  rescue boats}.
\newblock \bibinfo{journal}{European Journal of Operational Research}
  \bibinfo{year}{2016};\bibinfo{volume}{254}:\bibinfo{pages}{279--293}.
\newblock \DOIprefix\doi{10.1016/j.ejor.2016.03.026}.
%Type = Book
\bibitem[{{U.S. Coast Guard}(2013)}]{NatSAR}
\bibinfo{author}{{U.S. Coast Guard}\xfnm[]}.
\newblock \bibinfo{title}{The U.S. coast guard addendum to the united states
  national search and rescue supplement to the international aeronautical and
  maritime search and rescue manual}.
\newblock \bibinfo{publisher}{COMDTINST M16130.2F}, \bibinfo{year}{2013}.
%Type = Article
\bibitem[{{U.S. Coast Guard}(2014)}]{SARPlan}
\bibinfo{author}{{U.S. Coast Guard}\xfnm[]}.
\newblock \bibinfo{title}{Fourteenth coast guard district search and rescue
  plan}.
\newblock \bibinfo{journal}{CGD14INST M161301A} \bibinfo{year}{2014};.
%Type = Article
\bibitem[{Zemel(1984)}]{Zemel1984}
\bibinfo{author}{Zemel\xfnm[ E.]}.
\newblock \bibinfo{title}{Probabilistic analysis of geometric location
  problems}.
\newblock \bibinfo{journal}{Annals of Operations Research}
  \bibinfo{year}{1984};\bibinfo{volume}{1}:\bibinfo{pages}{215--238}.

\end{thebibliography}

\end{document}